# Exact Multistate Reliability and Upgrade Design for Heterogeneous HBM Systems via Threshold-Pruned BAT

**Wei-Chang Yeh***
Department of Industrial Engineering and Engineering Management, National Tsing Hua University, Hsinchu 300044, Taiwan
* Correspondence: yeh@ieee.org

**Abstract –** High Bandwidth Memory (HBM) systems can exhibit partial service rather than only full service or complete isolation: a controller-visible service unit may deliver full, reduced, or zero bandwidth because of sub-channel isolation, lane remapping, or protection overhead. This paper develops an exact multistate reliability framework in which each service unit carries an arbitrary finite set of bandwidth states and reliability is the probability that aggregate delivered bandwidth meets a demand. The binary k-out-of-n model is recovered as a special case, while closed-form binary-mapping error relations quantify mean-bandwidth distortion and provide a screening test for the simpler abstraction. For exact single-threshold evaluation a threshold-pruned multistate Binary-Addition-Tree (TP-mBAT) algorithm is proposed. It is deliberately regime-specific: fixed-grid dynamic programming is preferable on a compact common grid, where a 16-unit commensurate control required 273 pruned-DP updates versus 362,506 TP-mBAT node visits. On a reproducible 14-unit incommensurate benchmark, TP-mBAT is compared against a dynamic program carrying the same threshold rules, so that no baseline is weakened. Both expand the same state space, 6,862 nodes against 6,861 updates, and the separation lies entirely in retained state: 17 traversal entries against 412,121 probability states at central demand, reducing measured peak storage from 25.02 MB to 1,152 B. An exact probability-transfer sensitivity identifies when moving mass from a degraded state to a higher-bandwidth state changes system success, and a reserved-unit floor model admits a third exact pruning rule that is vacuous without such floors. A latent package-state mixture captures shared stress, where ignoring dependence overstates reliability by 8.73 percentage points.

**Keywords:** High Bandwidth Memory; multistate reliability; bandwidth reliability; Binary-Addition-Tree; BAT; heterogeneous systems; incommensurate bandwidth; exact threshold evaluation; sensitivity analysis

## 1. INTRODUCTION

High Bandwidth Memory (HBM) has become a key technology for AI accelerators, high-performance computing, and bandwidth-intensive processors. Its three-dimensional integration provides high data rates but also creates complex interactions among DRAM, TSV, bonding interfaces, thermal effects, and controller-level management. The system-level reliability question is therefore not a binary alive/failed classification, but whether the remaining memory bandwidth can meet the workload requirement [1–9].

The binary abstraction treats each controller-visible service unit as contributing either full bandwidth or zero, which reduces reliability evaluation to a tractable threshold problem. This representation is appropriate when intermediate degraded states are negligible [1–9]. However, modern memory systems may exhibit intermediate service conditions: a sub-channel may be partially isolated while the remainder continues to serve requests; remapping or protection mechanisms may reduce effective bandwidth; and different units may have different nominal capacities or failure probabilities. Under these conditions, the system cannot be characterized solely by the number of operational binary components; the relevant random variable becomes the delivered bandwidth itself [22–25].

This paper builds a hierarchical multistate reliability framework at the controller-visible service-unit level, without assuming a one-to-one correspondence between physical DRAM dies and memory channels. Each unit may occupy an arbitrary finite set of bandwidth states. Unit capacities are aggregated into stacks and then into system-level bandwidth, and reliability is evaluated against a specified demand. Universal generating functions and dynamic aggregation provide exact distribution-based baselines [26–28], while the proposed threshold-pruned multistate binary-addition-tree method (TP-mBAT) evaluates a specified threshold directly through safe branch pruning. The companion binary study established the binary pooled-bandwidth formulation [29]; the present paper addresses the complementary regime where binary service states are insufficient.

TP-mBAT is designed for a specific computational regime, not as a universal replacement for dynamic programming. When bandwidth states lie on a compact common grid, fixed-grid DP is efficient; TP-mBAT targets heterogeneous and effectively incommensurate bandwidth values, where exact sparse DP may accumulate a large number of distinct partial sums. The pruning logic of TP-mBAT specialises the upper and lower feasibility bounds of GE-MBAT [17] to the additive bandwidth setting, where general max-flow computations reduce to scalar comparisons. When reserved service units provide positive guaranteed capacity floors, an additional exact pruning rule becomes available.

The main contributions are:

1. **TP-mBAT algorithm** – a threshold-pruned multistate BAT with traversal storage bounded by $O(U \cdot L_{\max})$.
2. **Solver-relationship theorem (Proposition 1)** – establishing the computational boundary between dynamic programming and TP-mBAT, providing a clear solver-selection rule.
3. **Probability-transfer sensitivity and heterogeneous upgrade design** – identifying the residual-capacity interval in which a state upgrade changes system success, and demonstrating minimum-cost heterogeneous unit upgrades.
4. **Extensions for shared dependence and reserved-unit floors** – a latent package-state mixture captures common-cause dependence, and reserved-unit floors introduce a third exact pruning rule based on the positive guaranteed contribution of unassigned reserved units.

When each service unit is restricted to full-bandwidth or failed states, the multistate formulation reduces exactly to the binary pooled-bandwidth model [29]. This paper focuses on the broader regime where intermediate service levels, heterogeneous capacities, and non-negligible degraded-state probabilities must be retained explicitly.

## 2. MULTISTATE HBM SERVICE MODEL

A memory controller exposes service units and their service levels to a workload, not physical dies or fault states. The model built here takes this exposed view as its primitive: controller-visible service units carry multistate bandwidth variables; units aggregate into stacks and stacks into systems; pooled and replicated service are kept distinct because they are different reliability events; and a latent package state captures dependence arising from shared package-level or operating conditions.

### 2.1. Controller-Visible Service Unit and Bandwidth States

Consider $S$ HBM stacks; stack $s$ contains $C_s$ controller-visible service units, with total unit count $U = \sum_s C_s$. Unit $u$ has a random bandwidth contribution $X_u$ taking values from the finite set $\mathcal{A}_u = \{a_{u,0}, \ldots, a_{u,L_u}\}$, where $a_{u,0} = 0$ is the failed state and $a_{u,L_u}$ is the full bandwidth. The associated probabilities are $p_{u,l}$:

$$X_u \in \mathcal{A}_u,\ 0=a_{u,0}<\cdots<a_{u,L_u}, \tag{1}$$

$$\Pr(X_u = a_{u,l}) = p_{u,l},\ \sum_{l=0}^{L_u} p_{u,l} = 1. \tag{2}$$

The three-state case $\{0, \alpha_u, b_u b_u\}$ is useful, but no derivation below relies on it. Stack bandwidth is $W_s = \sum_{i=1}^{C_s} X_{s,i}$, and total system bandwidth is $W_{\text{tot}} = \sum_u X_u$. Pooled reliability is the probability that the sum of all unit capacities meets demand $D$:

$$R_{\text{pool}}(D) = \Pr(W_{\text{tot}} \geq D). \quad (3)$$

When multiple stacks are alternatives rather than additive bandwidth contributors, a different event applies. If success means at least one stack reaches a local threshold $\tau$, the reliability is $1 - [1 - R_{\text{stack}}(\tau)]^S$. This formula is not applicable to pooled bandwidth; the correct event for pooling is Equation (3), which depends on the distribution of the sum. This distinction is central to the framework.

**2.2. Conditional Dependence through a Shared Package State**

Service-unit states may be correlated when they share thermal, power, manufacturing, package-level, or other common conditions. Introduce a discrete latent state $H \in \{1, \ldots, K\}$ with $\pi_h = \Pr(H = h)$. Conditional on $H = h$, units are independent with state probabilities $p_{u,l|h}$. The unconditional pooled reliability is a finite mixture of conditional reliabilities:

$$R_{\text{pool}}(D) = \sum_{h=1}^{K} \pi_h \ R_{\text{pool}}(D \mid H = h). \quad (4)$$

Stack-specific latent states can be introduced analogously when stacks experience distinct regimes. This construction can induce positive dependence and also provides a calibration path when field data classify operating conditions into a small number of package regimes.

**2.3. Reserved Service Units and Capacity Floors**

Not every unit is interchangeable with its neighbours. Some channels may be pinned to latency-critical traffic, reserved for metadata or error-correction overhead, or bound to a compute tile. In such cases, surplus bandwidth elsewhere cannot substitute for that unit's contribution, so aggregate capacity alone may overstate deliverable service.

Let $\mathcal{R} \subseteq \{1, \ldots, U\}$ be the set of reserved units, and let $c_u > 0$ be the capacity floor that unit $u$ must individually meet. The reliability of interest is the joint event that the pooled demand is met and every reserved unit clears its floor:

$$R_{\text{res}}(D) = \Pr(\textstyle\sum_{u=1}^{U} X_u , \geq, D \ X_u \geq c_u \ \forall u \in \mathcal{R}). \quad (5)$$

Two limits are worth naming: $\mathcal{R} = \emptyset$ recovers the pooled model; $\mathcal{R} = \{1, \ldots, U\}$ gives a fully reserved system, though the pooled demand remains an additional requirement unless the floors jointly imply it.

The constraint is enforced at the state-set level rather than by a post hoc filter. For each reserved unit, every state with bandwidth below $c_u$ is deleted from $\mathcal{A}_u$. The retained probabilities are not renormalized: the removed mass corresponds to floor violations, which are failure events and must not be redistributed. Feasibility requires $c_u \leq a_{u,L_u}$ for every $u \in \mathcal{R}$; otherwise $R_{\text{res}}(D) = 0$. After deletion, the smallest retained state at or above $c_u$ defines the effective floor:

$$\hat{c}_u = \min\{a_{u,l} : a_{u,l} \geq c_u\}, u \in \mathcal{R}. \quad (6)$$

This satisfies $\hat{c}_u \geq c_u$, providing positive lower bounds on the contributions of unassigned reserved units. Section 4.5 uses these bounds to derive an additional exact pruning rule. When $\mathcal{R} = \emptyset$, the rule is inactive and the model reduces to the unreserved case.

## 3. EXACT BANDWIDTH RELIABILITY

Exact bandwidth reliability follows from the aggregate capacity distribution, with the binary model recovered as a special case.

### 3.1. Probability Generating Function and Dynamic Aggregation

Assume all bandwidth values are integer multiples of a basic increment $\Delta$, and let $d_{u,l} = a_{u,l}/\Delta$. The probability generating function of unit $u$ is

$$G_u(z) = \sum_{l=0}^{L_u} p_{u,l}\, z^{d_{u,l}}. \quad (7)$$

If units are unconditionally independent, the total bandwidth generating function is the product $G_{\text{tot}}(z) = \prod_u G_u(z)$. The coefficient of $z^d$ gives the exact probability that total bandwidth equals $d\Delta$. Let $d^* = \lceil D/\Delta \rceil$; then the pooled reliability is

$$R_{\text{pool}}(D) = \sum_{d \geq d^*} [z^d] G_{\text{tot}}(z). \quad (8)$$

When the latent dependence structure of Section 2.2 is present, the product form applies conditionally on $H = h$. Define the conditional generating functions $G_{u|h}(z) = \sum_l p_{u,l|h}\, z^{d_{u,l}}$ and $G_{\text{tot}|h}(z) = \prod_u G_{u|h}(z)$. The unconditional pooled reliability is then

$$R_{\text{pool}}(D) = \sum_{h=1}^{K} \pi_h \sum_{d \geq d^*} [z^d] G_{\text{tot}|h}(z). \quad (9)$$

When $K = 1$, this reduces to the unconditional-independent case.

Equations (8)–(9) can be evaluated by repeated convolution. If bandwidth values lie on a compact common grid, a fixed-grid DP stores probabilities by integer capacity index. Let $M_{\max} = \sum_u d_{u,L_u}$ be the maximum discrete aggregate-capacity index, and let $f_r(d)$ be the probability that the first $r$ units deliver total discrete capacity $d$. With $f_0(0) = 1$, the recursion is

$$f_r(d) = \sum_{l=0}^{L_r} p_{r,l}\, f_{r-1}(d - d_{r,l}). \tag{10}$$

This recursion computes the full distribution exactly with time $O(UL_{\max}M_{\max})$ and memory $O(M_{\max})$. When bandwidths are heterogeneous and admit only a very fine common integer scaling, the grid may become impractically large; a sparse exact DP can retain only reachable sums (represented as scaled integers, rationals, or another collision-free exact representation), but its dictionary size can approach $\prod_u (L_u + 1)$. This distinction motivates TP-mBAT: it avoids building the complete reachable-sum set when only a single threshold is required and threshold pruning is effective.

### 3.2. Binary $k$-out-of-$n$ as a Special Case

If all $U$ units are identical and have only two states $\{0, b\}$ with full-service probability $p$, then $G_u(z) = 1 - p + pz$ (taking $\Delta = b$), and the total generating function is $(1 - p + pz)^U$. For demand $D = kb$, the coefficient expansion gives

$$R_{\text{pool}}(kb) = \sum_{j=k}^{U} \binom{U}{j} p^j (1-p)^{U-j}. \tag{11}$$

Thus, the classical $k$-out-of-$n$ expression (with $n = U$) is exactly recovered. The multistate formulation is therefore a strict generalization, not a competing model. Equation (11) is the classical binary k-out-of-n reliability formula [18–21], obtained here as the two-state restriction of the multistate model rather than assumed.

### 3.3. Expected-Bandwidth Error Induced by Binary Mapping

For a three-state unit $X \in \{0, \alpha, bb\}$ with probabilities $p_{\text{fail}}, p_{\text{deg}}, p_{\text{full}}$, the exact mean bandwidth is

$$\mu = b(p_{\text{full}} + \alpha p_{\text{deg}}). \tag{12}$$

An optimistic binary mapping replaces the degraded state $\alpha b$ by $b$, while a conservative mapping replaces it by zero. For $\mu > 0$, the corresponding relative mean-bandwidth errors are

$$\varepsilon_{\text{opt}} = \frac{(1-\alpha)p_{\text{deg}}}{p_{\text{full}} + \alpha p_{\text{deg}}}, \tag{13}$$

$$\varepsilon_{\text{cons}} = \frac{\alpha p_{\text{deg}}}{p_{\text{full}} + \alpha p_{\text{deg}}}. \tag{14}$$

These expressions generalize the half-bandwidth relation of Paper 1 [29, Eqs. (33)–(34)] to arbitrary $\alpha$. At $\alpha = 0.5$, the two errors coincide at $p_{\text{deg}}/(2p_{\text{full}} + p_{\text{deg}})$; for $\alpha \neq 0.5$, they are asymmetric. These quantities measure expected-bandwidth distortion only; threshold-reliability error is examined separately in Section 6.2.

## 4. THRESHOLD-PRUNED MULTISTATE BAT

Exact threshold evaluation is #P-hard in general. TP-mBAT therefore prunes subtrees whenever threshold bounds certify success or failure, targeting heterogeneous or effectively incommensurate single-threshold problems while leaving compact-grid cases to dynamic programming.

### 4.1. State Representation, Pruning Rules, Algorithm, and Correctness

Let $x_u \in \{0, 1, \ldots, L_u\}$ denote the selected state index of service unit $u$. A complete system state is $x = (x_1, \ldots, x_U)$. Its bandwidth and probability are

$$B(x) = \sum_u a_{u,x_u}, \Pr(x) = \prod_u p_{u,x_u}. \tag{15}$$

The correctness of TP-mBAT does not depend on service-unit order. For reproducibility, the numerical study orders units by decreasing maximum bandwidth contribution $a_{u,L_u}$. This is a deterministic heuristic rather than a claim of optimal ordering; alternative orderings can change the number of visited nodes without changing the returned probability.

At a partial state after $r$ units have been assigned, let $B_r$ be the accumulated bandwidth, $P_r$ the probability of the partial assignment, and $M_{\text{rem}}(r)$ the sum of the maximum possible bandwidths of all unassigned units:

$$M_{\text{rem}}(r) = \sum_{u=r+1}^{U} a_{u,L_u}. \tag{16}$$

Two exact rules follow.

**Success aggregation rule.** If $B_r \geq D$, every descendant state meets the demand. Because the probabilities of all remaining states sum to one, the entire descendant probability mass equals $P_r$ and can be added to reliability immediately.

**Failure pruning rule.** If $B_r + M_{\text{rem}}(r) < D$, even the maximum possible contribution of all remaining units cannot meet the demand. The entire branch contributes zero and is discarded.

Both rules are the additive specialisation of the $F^+$ and $F^-$ bounds of GE-MBAT [17], which are themselves an extension of the infeasible-vector tests of QBAT [11]. Section 4.4 states the correspondence and its consequences for what is and is not claimed as new here.

Algorithm 1 states the procedure for computing $R_{\text{pool}}(D)$. Units are processed in the order fixed above, and the two rules are applied at every node before any child is generated, so that no state excluded by either rule is ever created. For $R_{\text{res}}(D)$, the modifications described in Section 4.5 apply.

**Algorithm 1.** TP-mBAT for $R_{\text{pool}}(D)$

Input: state values $a[u, l]$, probabilities $p[u, l]$, demand $D$

Output: exact pooled-bandwidth reliability $R$

1. Reorder service units, default: decreasing maximum capacity
2. Precompute M_rem[r] from Equation (16)
3. R ← 0
4. Push root node (r=0, B=0, P=1) onto a stack
5. while stack is not empty:
6. Pop (r, B, P)
7. if B ≥ D:
8. R ← R + P
9. continue
10. if B + M_rem[r] < D:
11. continue
12. if r = U:
13. continue
14. u ← r + 1
15. for each state l of service unit u, in decreasing capacity:
16. Push (r+1, B+a[u,l], P·p[u,l])
17. return R

For the latent dependence model of Equation (4), TP-mBAT is executed independently under each package state $h$ using $p_{u,l|h}$. The final reliability is the $\pi_h$-weighted sum. The conditional calculations are independent and can be parallelized.

**Theorem 1.** TP-mBAT returns the exact value of $R_{\text{pool}}(D)$ under the state probabilities supplied to the algorithm.

**Proof.** Each complete multistate vector appears exactly once in the underlying mixed-radix addition tree. A branch satisfying $B_r \geq D$ contains only successful descendants, so replacing those descendants by their total conditional probability of one preserves their combined mass $P_r$. A branch satisfying $B_r + M_{\text{rem}}(r) < D$ contains only failing descendants and contributes zero. Every branch not removed by either rule is expanded until one of the rules applies or a terminal state is reached. Therefore no successful probability mass is omitted or duplicated, and the returned sum equals $\Pr(W_{\text{tot}} \geq D)$. □

### 4.2. Worked Example

A three-unit instance is small enough to trace completely and is chosen so that both pruning rules fire. The service units are heterogeneous and their capacities admit no common grid coarser than $0.1b_0$.

**Table 1.** Worked example: three heterogeneous three-state service units.

| Unit | Full | Degraded | Failed |
|---|---|---|---|
| $u1$ | $1.0b0$ (0.70) | $0.6b0$ (0.20) | 0 (0.10) |
| $u2$ | $0.8b0$ (0.60) | $0.5b0$ (0.30) | 0 (0.10) |
| $u3$ | $0.7b0$ (0.50) | $0.3b0$ (0.30) | 0 (0.20) |

The demand is $D = 1.5b_0$. Units are ordered by decreasing maximum capacity, which is already the listed order, so the maximum attainable remainders are $M_{\text{rem}}(0) = 2.5$, $M_{\text{rem}}(1) = 1.5$, $M_{\text{rem}}(2) = 0.7$, and $M_{\text{rem}}(3) = 0$.

Children are pushed in decreasing capacity, so the lowest-capacity child is popped first. This affects peak stack occupancy but not reliability or node count.

TP-mBAT visits 25 nodes, eight of which require expansion, and agrees exactly with exhaustive enumeration. The threshold-pruned DP likewise performs eight surviving-state expansions, as predicted by Proposition 1 for a collision-free surviving set. The complete 25-node traversal is given in Supplementary Table S1.

### 4.3. Computational Complexity and Solver Relationship

**Remark 1 (computational hardness).** Exact evaluation of $\Pr(\sum_u X_u \geq D)$ is #P-hard. Restricting every unit to two states $\{0, a_u\}$ with equal probabilities gives

$$\Pr(\textstyle\sum_u X_u \geq D) \cdot 2^U = |\{S \subseteq [U]: \sum_{u\in S} a_u \geq D\}| \tag{17}$$

and the difference between the values at $D$ and $D + 1$ counts the subsets of exact sum:

$$[\Pr(\textstyle\sum_u X_u \geq D) - \Pr(\sum_u X_u \geq D + 1)] \cdot 2^U = |\{S \subseteq [U]: \sum_{u\in S} a_u = D\}| \tag{18}$$

Thus #SUBSET-SUM reduces to exact threshold reliability [30,31], establishing #P-hardness and exponential worst-case behavior.

Compact integer grids nevertheless admit pseudo-polynomial dynamic programming, whereas exact scaling of effectively incommensurate capacities can require an impractically fine grid. Difficulty depends on both capacity collisions and how surviving partial sums cluster near the threshold, as examined in Section 6.3.

Because the objective is an additive sum, the same success and infeasibility bounds are available to dynamic programming. The fair comparator is therefore a threshold-pruned DP using the same rules.

It follows that the correct comparator is a dynamic program applying the same two rules, not an unpruned one. Let the surviving set after $r$ assignments be

$$\Sigma_r = \{\text{partial assignments}, x_1, \ldots x_r : B_r < D \text{ and } B_r + M_{\text{rem}}(r) \geq D\}, \tag{19}$$

that is, the partial assignments that are neither already successful nor already infeasible. The condition $|C_r| = |\Sigma_r|$ at every $r$ corresponds to the case where the accumulated capacities are collision-free; a superincreasing capacity set is an extreme example.

**Proposition 1 (solver relationship).** Let a threshold-pruned dynamic program apply the success-collapse and infeasibility-pruning rules of Section 4.1 under the same unit ordering as TP-mBAT, and write $C_r = \{B(x) : x \in \Sigma_r\}$ for the set of distinct accumulated capacities among the surviving partial assignments at position $r$. Then the dynamic program retains and expands exactly one state for each element of $C_r$, aggregating the probability masses of assignments that share a capacity, while TP-mBAT expands every assignment in $\Sigma_r$ individually. Consequently the dynamic-program expansion count never exceeds that of TP-mBAT, with equality when the surviving capacities are collision-free, that is when $|C_r| = |\Sigma_r|$ at every $r$. The peak traversal storage of TP-mBAT is $O(U \cdot L_{\max})$ irrespective of $|\Sigma_r|$, whereas the dynamic program retains $|C_r|$ probability states.

**Proof.** Both methods discard a partial assignment exactly when $B_r \geq D$, in which case the whole descendant mass $P_r$ is added, or when $B_r + M_{\text{rem}}(r) < D$, in which case the branch contributes zero. The surviving sets are therefore identical. The dynamic program additionally combines states with equal accumulated capacity, which can only reduce the number of subsequent expansions, giving the inequality; when capacities are collision-free no such combination occurs and the counts coincide. TP-mBAT stores only the current root-to-node path and its pending siblings. With all children pushed onto the stack at each expanded node, a tighter bound is $1 + \sum_{u=1}^{U} L_u \leq 1 + U \cdot L_{\max}$, so the stack is bounded by $O(U \cdot L_{\max})$ regardless of how many assignments survive. □

Proposition 1 concerns expansions rather than time: pruned DP never expands more surviving states and gains whenever equal capacities can be merged. TP-mBAT instead offers bounded traversal storage, which matters when the surviving set is wide; Sections 6.3 and 6.7 report the corresponding experiments.

### 4.4. Relation to GE-MBAT and Scope of Contribution

GE-MBAT [17] uses upper and lower completion bounds by assigning unassigned arcs to full capacity or zero, with the underlying tests traced to QBAT [11]. In the additive HBM setting, these max-flow bounds reduce to scalar capacity comparisons at each node. The mixed-radix encoding is inherited from multistate BAT [12], and subtree probability aggregation follows the super-vector concept of bounded and quick BAT variants [15]. These inherited elements are not claimed as new.

The contribution is their HBM-specific characterization: scalar-bound evaluation, the traversal-memory guarantee, Proposition 1's exact relationship to pruned DP, the resulting solver-selection boundary, and the reserved-floor rule of Section 4.5. Proposition 1 is central; TP-mBAT is not claimed to be universally faster.

Table 2 summarizes the resulting comparison with GE-MBAT.

**Table 2.** Comparison with GE-MBAT [17].

| Aspect | GE-MBAT [17] | Proposed evaluator |
|---|---|---|
| System | Multistate flow network with topology | Unordered service units, no topology |
| Success criterion | Maximum flow $F(X) \geq d$ | Coordinate sum $\sum u\ Xu \geq D$ |
| Upper bound | $F^{+}$: max-flow with unassigned arcs at full capacity | $B_{r} + M_{rem}(r)$, a running sum |
| Lower bound | $F^{-}$: max-flow with unassigned arcs at zero | $B_{r}$, a running sum |
| Cost per bound | Maximum-flow evaluation | One addition and one comparison |
| Application of bounds | At selected points, with layered-arc decomposition to limit the number of flow computations | At every node; no decomposition required |
| Guaranteed-success rule | Vacuous: $F^{-}$ sets unassigned arcs to zero | Active under the floors of Section 2.3 |
| Constraint class | Global demand level only | Global demand plus per-unit floors on a subset |

### 4.5. An Additional Exact Rule under Reserved-Unit Floors

The two rules of Section 4.1 use the accumulated bandwidth $B_r$ and the maximum attainable remainder $M_{\text{rem}}(r)$. Under the reserved-unit model of Section 2.3 a third quantity becomes available: a guaranteed minimum remainder, formed from the effective floors of the reserved units not yet assigned. Define the retained-mass product for the unassigned reserved units:

$$\Pi_{\text{rem}}(r) = \prod_{\substack{u>r \\ u\in\mathcal{R}}} \sum\nolimits_{l:a_{u,l}\geq c_u} p_{u,l}. \tag{20}$$

Using the effective floor $\hat{c}_u$ rather than the floor $c_u$ itself is free: both are known once the sub-floor states have been deleted, but the effective floor is never smaller and is usually strictly larger. The guaranteed minimum remainder is

$$m_{\text{rem}}(r) = \sum_{\substack{u>r \\ u\in\mathcal{R}}} \hat{c}_u. \tag{21}$$

In the unreserved case $\mathcal{R}$ is empty, $m_{\text{rem}}(r)$ is identically zero, and the rule below can never fire. It is therefore not an improvement to the unreserved algorithm but a rule that the floor constraints themselves create.

**Guaranteed-success rule.** If $B_r + m_{\text{rem}}(r) \geq D$, then every descendant of the node that satisfies all remaining floors also meets the demand, so the branch closes immediately and contributes $P_r \cdot \Pi_{\text{rem}}(r)$.

**Theorem 2.** With the success aggregation rule of Section 4.1 restated so that a closing node contributes $P_r \cdot \Pi_{\text{rem}}(r)$, the addition of the guaranteed-success rule leaves the returned value unchanged and equal to $R_{\text{res}}(D)$.

**Proof.** Consider a node closed by the new rule at position $r$. Its descendants that violate some remaining floor are excluded from the event by construction and their mass has already been removed from the retained state sets, so the descendant mass that remains in the event is exactly $P_r \cdot \Pi_{\text{rem}}(r)$. Among those descendants, each assigns every remaining reserved unit one of its retained states, and every retained state is at least that unit's effective floor, hence a total of at least $B_r + m_{\text{rem}}(r) \geq D$, so all of them are successes. Replacing the subtree by $P_r \cdot \Pi_{\text{rem}}(r)$ therefore neither omits nor duplicates any successful mass. The same weighting is required of the original success rule: on a reserved instance the deleted states are floor violations rather than absent outcomes, so a node closed because $B_r \geq D$ also contributes only $P_r \cdot \Pi_{\text{rem}}(r)$, and the unweighted form would over-count. When $\mathcal{R}$ is empty, $\Pi_{\text{rem}}(r) = 1$, $m_{\text{rem}}(r) = 0$, and both statements reduce to those of Theorem 1. □

The rule changes the number of nodes expanded but not the storage bound: it closes subtrees earlier without adding anything to the traversal stack, so the $O(U \cdot L_{\max})$ bound of Proposition 1 is unaffected.

Closing a subtree earlier changes floating-point accumulation order. Across the reservation levels of Section 6.8, the largest difference between the two- and three-rule variants is $6.33 \times 10^{-15}$; the rule is exact in exact arithmetic.

Unlike GE-MBAT's zero assignment for unassigned arcs, reserved-unit floors provide a positive lower bound on the contributions of unassigned reserved units. This positive guaranteed remainder enables the additional success rule. Without reserved-unit floors, the remainder is zero and the rule is vacuous.

## 5. RELIABILITY SENSITIVITY AND UPGRADE DESIGN

Exact reliability also yields probability-transfer sensitivities for heterogeneous upgrade design and screening of binary approximations.

### 5.1. Probability-Transfer Sensitivity and Heterogeneous Upgrades

Consider service unit $u$, transferring probability $\varepsilon$ from state $a$ with bandwidth $w_a$ to state $b$ with bandwidth $w_b > w_a$. Let $W_{-u} = \sum_{v \neq u} X_v$. Then

$$\frac{dR_{\text{pool}}}{d\varepsilon} = \Pr(D - w_b \leq W_{-u} < D - w_a). \tag{22}$$

Because reliability is linear in the state-probability vector of a single unit, for any feasible $0 \leq \varepsilon \leq p_{u,a}$, the exact change is

$$\Delta R_{\text{pool}} = \varepsilon \cdot \Pr(D - w_b \leq W_{-u} < D - w_a). \tag{23}$$

For the latent-mixture model, with conditional transfer $\varepsilon_h$,

$$\Delta R_{\text{pool}} = \sum_{h=1}^{K} \pi_h \, \varepsilon_h \cdot \Pr(D - w_b \leq W_{-u} < D - w_a \mid H = h). \tag{24}$$

This sensitivity naturally leads to a heterogeneous upgrade problem. For each unit $u$, an action $m \in \mathcal{M}_u$ has cost $c_{u,m}$ and replaces the state probabilities by $p_{u,l}(m)$. The minimum-cost reliability design is:

y* = arg min_y Σ_{u,m} c[u,m] y[u,m], s.t. R_pool(D; y) ≥ R_tar, Σ_m y[u,m] = 1 ∀u. (25)

Every action set contains a null action of zero cost that leaves the state probabilities unchanged, so selecting exactly one action per unit is equivalent to selecting a subset of units to upgrade.

Equation (25) can assign different actions to different units; solver choice follows Proposition 1 and the grid/collision structure of Sections 3.1 and 4.3. Under latent dependence, actions may specify conditional probabilities $p_{u,l|h}(m)$.

### 5.2. Mean-Bandwidth Screening for Binary Approximation

Equations (13)–(14) provide a fast screening test on expected bandwidth. Given tolerance $\epsilon$, the optimistic and conservative mapping errors satisfy respectively

$$\frac{(1-\alpha)p_{\text{deg}}}{p_{\text{full}}+\alpha p_{\text{deg}}} \leq \epsilon, \frac{\alpha p_{\text{deg}}}{p_{\text{full}}+\alpha p_{\text{deg}}} \leq \epsilon. \tag{26}$$

This test does not bound threshold-reliability error; counterexamples in Section 6.2 show that mean distortion and threshold error are not reliably ordered ($\alpha = 0.70$: 4.784% mean distortion vs. 8.042% threshold error; $\alpha = 0.30$: 5.109% mean distortion vs. 0% threshold error). The screening test is therefore only a filter on mean-bandwidth distortion: passing it does not license a binary model, and failing it does not by itself condemn one. Threshold decisions require direct evaluation of threshold error, which the exact solver of Section 4 provides.

## 6. NUMERICAL EXPERIMENTS

The experiments validate exactness, characterize solver-selection regimes, and quantify the principal modeling effects.

### 6.1. Experimental Setup, Baselines, and Validation

All timings were produced on a single host (Intel Core Ultra 7 265K, 20 physical cores) running Ubuntu 24.04, compiled with g++ 13.3.0 -O3 -std=c++17. Every solver is CPU-only and single-threaded; the process was pinned to a single logical CPU. Timing used a steady clock, reporting the median of 15 runs after 10 warm-up runs. Frequency scaling remained under OS control, so wall-clock times are implementation- and platform-dependent. Operation counts and peak-state counts are deterministic for fixed instances and reproducible on any platform. Reported reliabilities agree to double-precision numerical accuracy; validation experiments below show maximum absolute differences on the order of $10^{-15}$.

Algorithmic correctness was checked in two ways. First, 50 randomly generated three-state systems with $U = 4$ to 8 were evaluated by exhaustive enumeration and TP-mBAT; the maximum absolute probability difference was $3.22 \times 10^{-15}$. Second, 200 independently generated systems with $U = 4$ to 10 were compared between TP-mBAT and exact convolution; the maximum absolute difference was $7.66 \times 10^{-15}$. These discrepancies support Theorem 1 computationally.

All deterministic exact solvers use the same unit ordering and are compared under matched conditions; wall-clock time and byte allocation remain implementation-dependent.

Experiments cover the commensurate negative control, incommensurate solver benchmark, binary-mapping accuracy, service semantics, shared stress, Monte Carlo validation, heterogeneous upgrades, scalability, and reserved-unit floors. The primary comparator is a threshold-pruned sparse DP using the same two rules as TP-mBAT; naive and capped DPs, exhaustive enumeration, and Monte Carlo serve as secondary checks. Parameters are specified with each experiment.

### 6.2. Binary-Mapping Accuracy

Table 3 isolates the $\alpha \neq 0.5$ generalization. Eight identical three-state units use $(p_{\text{full}}, p_{\text{deg}}, p_{\text{fail}}) = (0.78, 0.14, 0.08)$ and $D = 6b$, with $\alpha$ varying away from 0.5. The $\alpha = 0.5$ sweep from Paper 1 [29] is not recomputed. Equations (13)–(14) predict unequal optimistic and conservative mean errors when $\alpha \neq 0.5$.

**Table 3.** Binary-mapping accuracy for eight units, $(p_{\text{full}}, p_{\text{deg}}, p_{\text{fail}}) = (0.78, 0.14, 0.08)$, $D = 6b$.

| α | Exact $R_{\text{pool}}$ | Mean $\varepsilon_{\text{opt}}$ (%) | Mean $\varepsilon_{\text{cons}}$ (%) | Threshold err. opt. (%) | Threshold err. cons. (%) |
|---|---|---|---|---|---|
| 0.30 | 0.751356 | 11.922 | 5.109 | +30.284 | 0.000 |
| 0.40 | 0.795721 | 10.048 | 6.699 | +23.020 | -5.576 |
| 0.60 | 0.883160 | 6.481 | 9.722 | +10.841 | -14.924 |
| 0.70 | 0.906039 | 4.784 | 11.162 | +8.042 | -17.072 |

As Figure 1 shows, the two binary mappings behave increasingly asymmetrically as the degraded-state bandwidth fraction moves away from the half-bandwidth case.

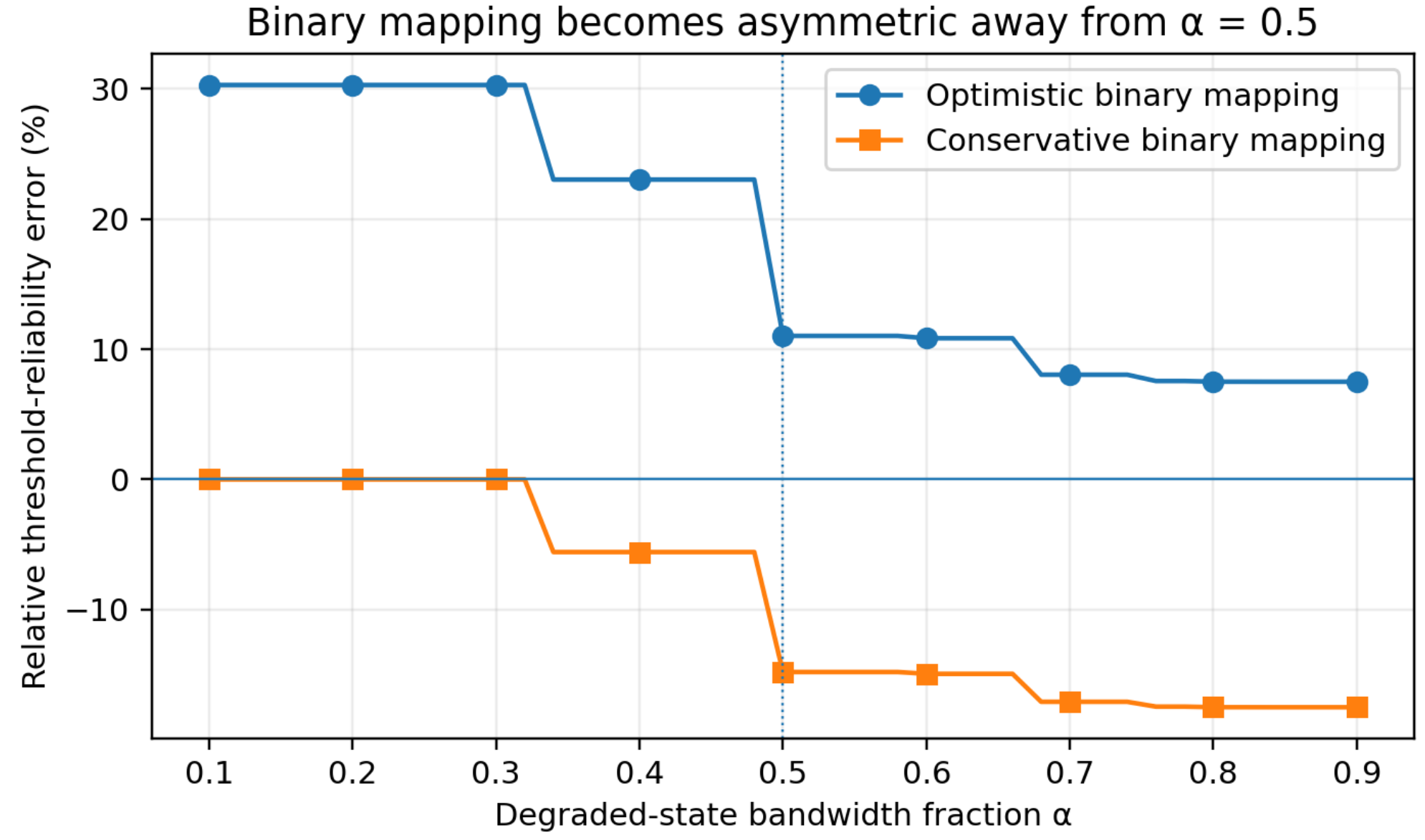

**Figure 1.** Relative threshold-reliability error as the degraded-state bandwidth fraction $\alpha$ varies for $U = 8$, $D = 6b$, and $(p_{\text{full}}, p_{\text{deg}}, p_{\text{fail}}) = (0.78, 0.14, 0.08)$. The half-bandwidth case $\alpha = 0.5$ is shown only as a reference line; its numerical sweep is reported in Paper 1 [29].

When degraded bandwidth is small, the conservative mapping can be exact because degraded states do not add enough capacity to change success, while the optimistic mapping overstates reliability by more than 30%. As $\alpha$ increases, the conservative mapping becomes pessimistic and the optimistic error shrinks. Thus the binary approximation depends on both $p_{\text{deg}}$ and the degraded-state service value $\alpha$.

### 6.3. Solver Selection: When TP-mBAT Helps and When It Does Not

Four exact solvers are compared: TP-mBAT; naïve sparse DP (no threshold rules); capped DP (success collapse only); and pruned DP (both rules). All return the same reliability to within $1 \times 10^{-12}$ Counting conventions: TP-mBAT operations are **node visits** (stack pops); DP operations are **candidate updates**. Peak states are maximum stack entries for TP-mBAT and maximum dictionary keys for the DPs.

For the collision-free benchmark rows reported below, the TP-mBAT node-visit count exceeds the pruned-DP candidate-update count by exactly one, corresponding to the TP root. This one-count offset follows from the reporting convention; Proposition 1 separately predicts equality of true expansion counts when surviving capacities are collision-free. The reported counters are consistent with this.

**Table 4.** Incommensurate benchmark at D = 0.875 of maximum aggregate capacity.

| U | Solver | Operations | Peak states | Peak bytes | Exact reliability |
|---|---|---|---|---|---|
| 10 | TP-mBAT | 658 | 5 | 288 B | 0.435493408392 |
| 10 | DP naïve | 88,572 | 59,049 | 2.799 MB | 0.435493408392 |
| 10 | DP capped | 88,569 | 58,983 | 2.797 MB | 0.435493408392 |
| 10 | DP pruned | 657 | 54 | 3,864 B | 0.435493408392 |
| 12 | TP-mBAT | 2,572 | 6 | 288 B | 0.425804070437 |
| 12 | DP naïve | 797,160 | 531,434 | 25.225 MB | 0.425804070437 |
| 12 | DP capped | 797,139 | 531,164 | 25.218 MB | 0.425804070437 |
| 12 | DP pruned | 2,571 | 230 | 15,096 B | 0.425804070437 |
| 14 | TP-mBAT | 6,862 | 7 | 288 B | 0.410855920925 |
| 14 | DP naïve | 7,173,300 | 4,780,410 | 227.429 MB | 0.410855920925 |
| 14 | DP capped | 7,173,258 | 4,779,738 | 227.413 MB | 0.410855920925 |
| 14 | DP pruned | 6,861 | 539 | 37,424 B | 0.410855920925 |

Against a naïve or capped DP, TP-mBAT appears up to roughly three orders of magnitude better in the reported operation counts, but that margin reflects a rule the DP is entitled to use. Against the pruned DP, the node-visit and update counts differ by exactly one; the separation is retained state: 7 stack entries versus 539 keys at $U = 14$.

**Table 5.** Peak-state and traversal-storage profile across demand, U = 14, incommensurate benchmark.

| $D$/max | TP-mBAT peak | DP pruned keys | Peak-state ratio | TP bytes | DP bytes | Storage ratio |
|---|---|---|---|---|---|---|
| 0.500 | 17 | 412,121 | 24,242× | 1,152 B | 25.02 MB | 21,718× |
| 0.650 | 13 | 166,010 | 12,770× | 576 B | 9.48 MB | 16,456× |

| *D*/max | TP-mBAT peak | DP pruned keys | Peak-state ratio | TP bytes | DP bytes | Storage ratio |
|---|---|---|---|---|---|---|
| 0.750 | 10 | 33,687 | 3,369× | 576 B | 2.00 MB | 3,469× |
| 0.875 | 7 | 539 | 77× | 288 B | 37.4 kB | 130× |
| 0.950 | 4 | 14 | 4× | 96 B | 1.24 kB | 13× |

Byte figures are measured allocation high-water marks and therefore include container and transient reallocation overhead; they are implementation-dependent and reported as empirical storage measurements. In this benchmark, the retained-state separation grows as the demand moves toward the central region of the attainable capacity range, where pruning is least effective. At $D = 0.50$, DP holds 412,121 states while TP-mBAT holds at most 17 entries. Figure 2 summarizes this retained-state behavior across system size and demand.

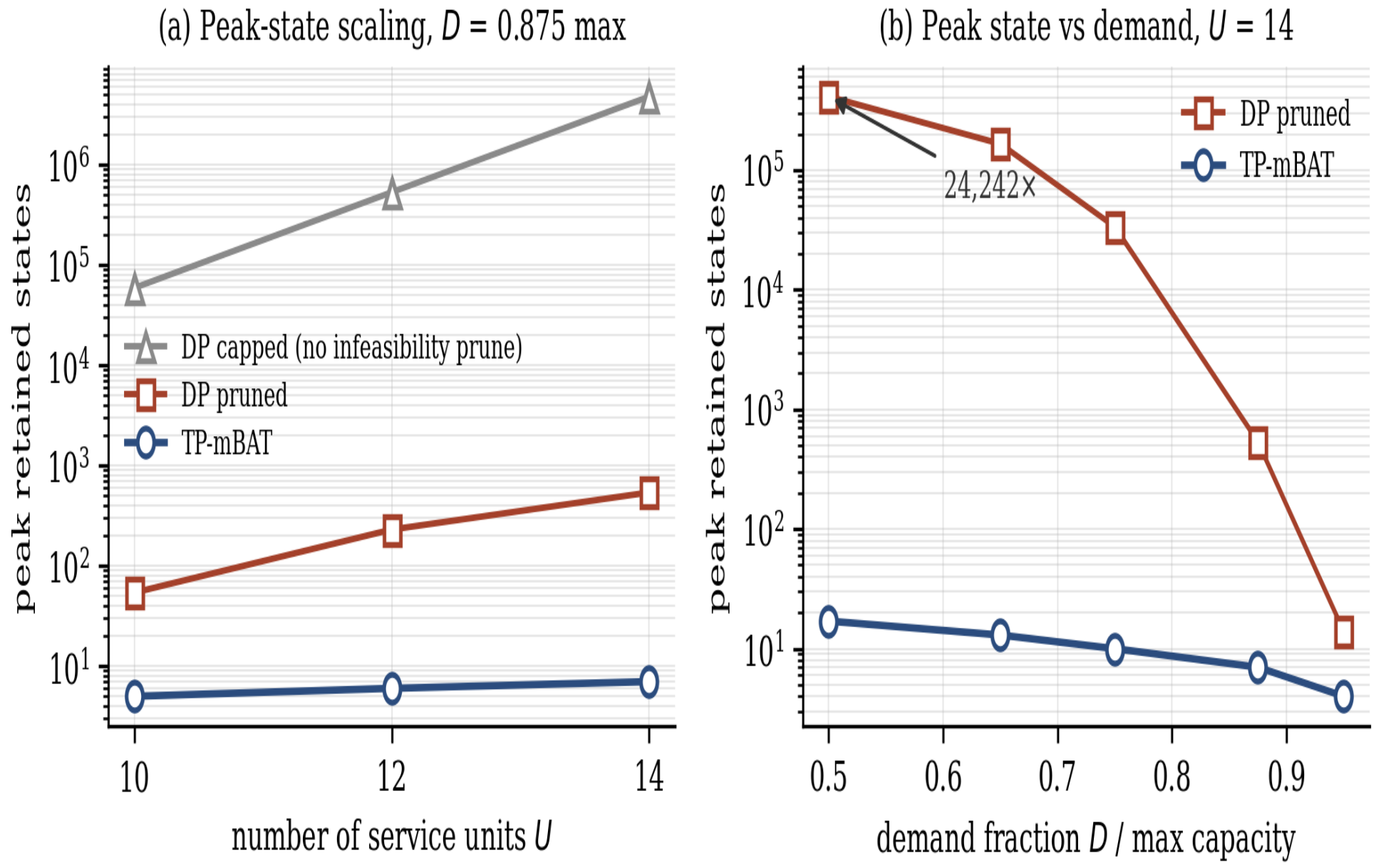

**Figure 2.** Peak retained states in the heterogeneous incommensurate benchmark. (a) Scaling with the number of service units at $D = 0.875$ of maximum capacity. (b) Dependence on demand fraction at $U = 14$. Both axes are logarithmic. The capped dynamic program is shown to isolate the effect of the infeasibility rule.

**Table 6.** Commensurate negative control, $U = 16$, states $\{0, 1, 2\}$, $D = 26$.

| Solver | Operations | Peak states | Exact reliability |
|---|---|---|---|
| TP-mBAT | 362,506 | 9 | 0.655673664947 |
| DP naive | 768 | 33 | 0.655673664947 |
| DP capped | 741 | 26 | 0.655673664947 |

| Solver | Operations | Peak states | Exact reliability |
|---|---|---|---|
| DP pruned | 273 | 7 | 0.655673664947 |

On a compact common grid, pruned DP is superior on both axes: 273 updates and 7 states against 362,506 node visits. Extensive capacity collisions allow DP to merge aggressively; the tree cannot exploit them. When a compact grid exists, DP is the correct solver.

**Table 7.** Effect of capacity structure at $U = 20$ binary units, $D = 0.5$ of maximum capacity. Collision-freeness alone does not make an instance hard. Peak states count retained probability states after each update and exclude the root, so a solver whose branches all close on their first test retains none; this is why the superincreasing instance reports zero.

| Capacity structure | TP-mBAT nodes | DP pruned updates | DP pruned peak states |
|---|---|---|---|
| Commensurate, small integer grid | 155,657 | 588 | 39 |
| Superincreasing ($3^i$), collision-free | 3 | 2 | 0 |
| Random large, unstructured | 163,703 | 163,702 | 18,802 |

The superincreasing row is instructive: capacities are collision-free, so DP cannot merge, yet both solvers terminate in a handful of operations because the bound is exact at every level. The hard regime is the random large, unstructured case, where neither merging nor bounding is effective; the node-visit and update counts differ only by the root, and the separation is retained state. Across 101 instances (seeds 20260812–20260912), TP-mBAT visits range from 146,935 to 351,499; the canonical instance is specified in Supplementary Section S4.2.

The experiments support a narrow selection rule: use fixed-grid or pruned DP when capacities share a compact common grid, when the full distribution is required, or when storage is not the constraint. Use TP-mBAT when capacities are effectively incommensurate, a single threshold is required, and storage is constrained: it carries out equivalent state-space exploration while traversal memory is bounded by $O(U \cdot L_{\max})$.

### 6.4. Service Semantics and Dependence Effects

This experiment verifies the distinction introduced in Section 2.1 between pooled bandwidth and replicated service. Each stack contains $C = 4$ units with $(p_{\text{full}}, p_{\text{deg}}, p_{\text{fail}}) = (0.70, 0.15, 0.15)$ and $\alpha = 0.5$. For comparison, the replicated local threshold is set to $\tau = D/S$. At high aggregate demand, the two events diverge sharply.

**Table 8.** Pooled-bandwidth and replicated-service reliability at high aggregate demand. Each stack holds $C = 4$ three-state units with $(p_{\text{full}}, p_{\text{deg}}, p_{\text{fail}}) = (0.70, 0.15, 0.15)$ and $\alpha = 0.5$.

| S | D/(SCb) | τ=D/S | $R_{rep}$ | $R_{pool}$ | Gap |
|---|---|---|---|---|---|
| 2 | 0.750 | 3b | 0.920391 | 0.687343 | +0.233048 |
| 3 | 0.875 | 3.5b | 0.829876 | 0.240830 | +0.589046 |
| 4 | 0.875 | 3.5b | 0.905735 | 0.185361 | +0.720374 |

Figure 3 illustrates the resulting divergence between replicated-service and pooled-bandwidth reliability at high aggregate demands.

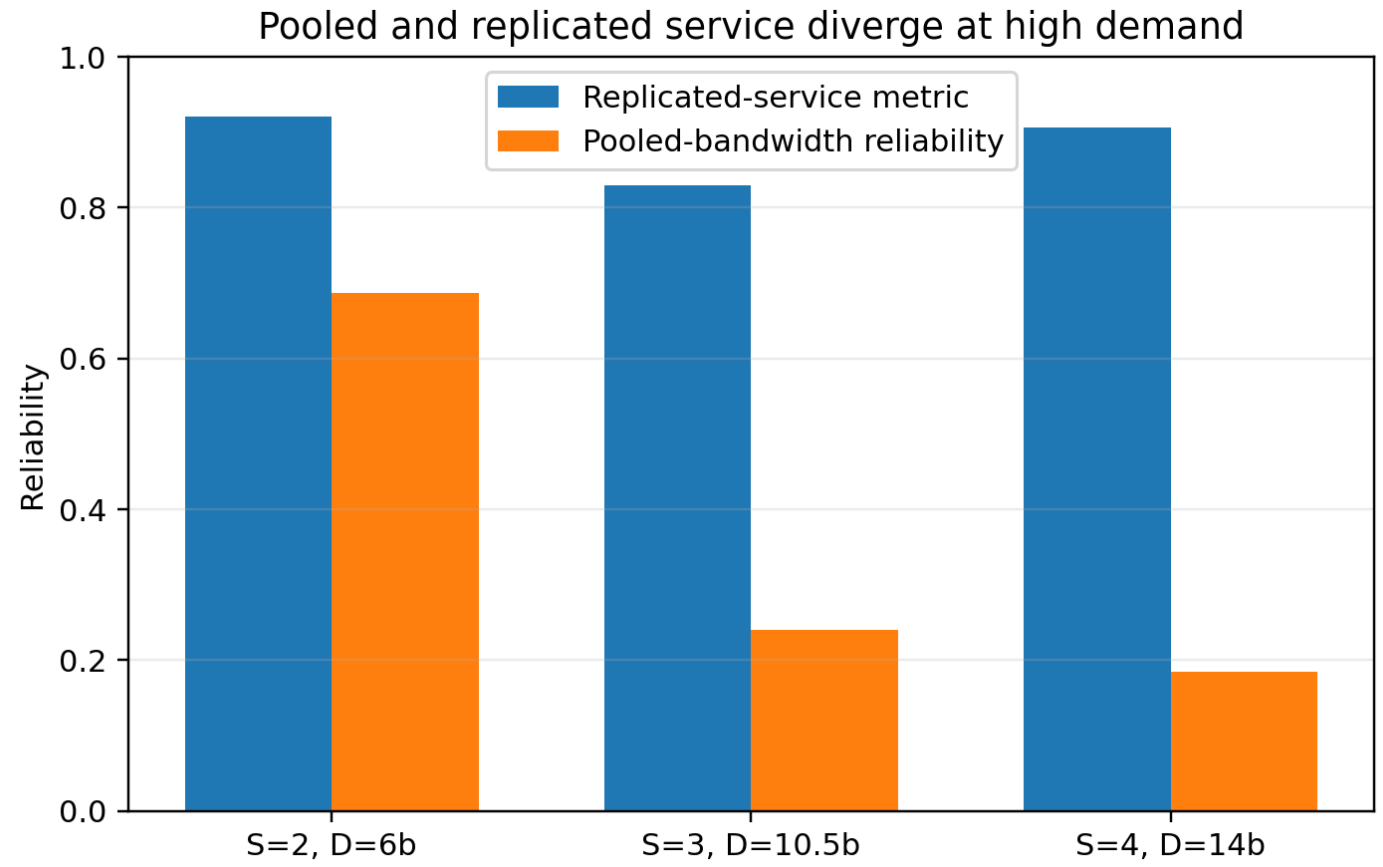

**Figure 3.** Replicated-service and pooled-bandwidth reliability for high aggregate demands. The two values answer different service questions and should not be substituted for one another.

This experiment evaluates the latent-mixture construction of Section 2.2 against an independence model with identical unconditional marginals. For $U = 16$ units with state values $\{0, 0.5, bb\}$, the normal conditional probabilities are $(0.85, 0.10, 0.05)$ and the stressed probabilities are $(0.55, 0.25, 0.20)$, with $D = 12b$. The comparator treats units as independent while preserving the same unconditional unit marginals.

**Table 9.** Effect of a shared latent package state for $U = 16$ and $D = 12b$, against an independence calculation built from the same unconditional unit marginals.

| P(stressed) | Exact latent mixture | Independent same marginals | Independence bias (pp) |
|---|---|---|---|
| 0.00 | 0.987389 | 0.987389 | 0.000 |
| 0.05 | 0.952150 | 0.979082 | +2.693 |
| 0.10 | 0.916911 | 0.967625 | +5.071 |
| 0.20 | 0.846433 | 0.933687 | +8.725 |

Figure 4 shows how the reliability bias caused by assuming independence increases with the probability of the stressed package state. At 20% stressed probability, independence overstates reliability by 8.725 percentage points. Unit marginals alone do not determine simultaneous capacity loss.

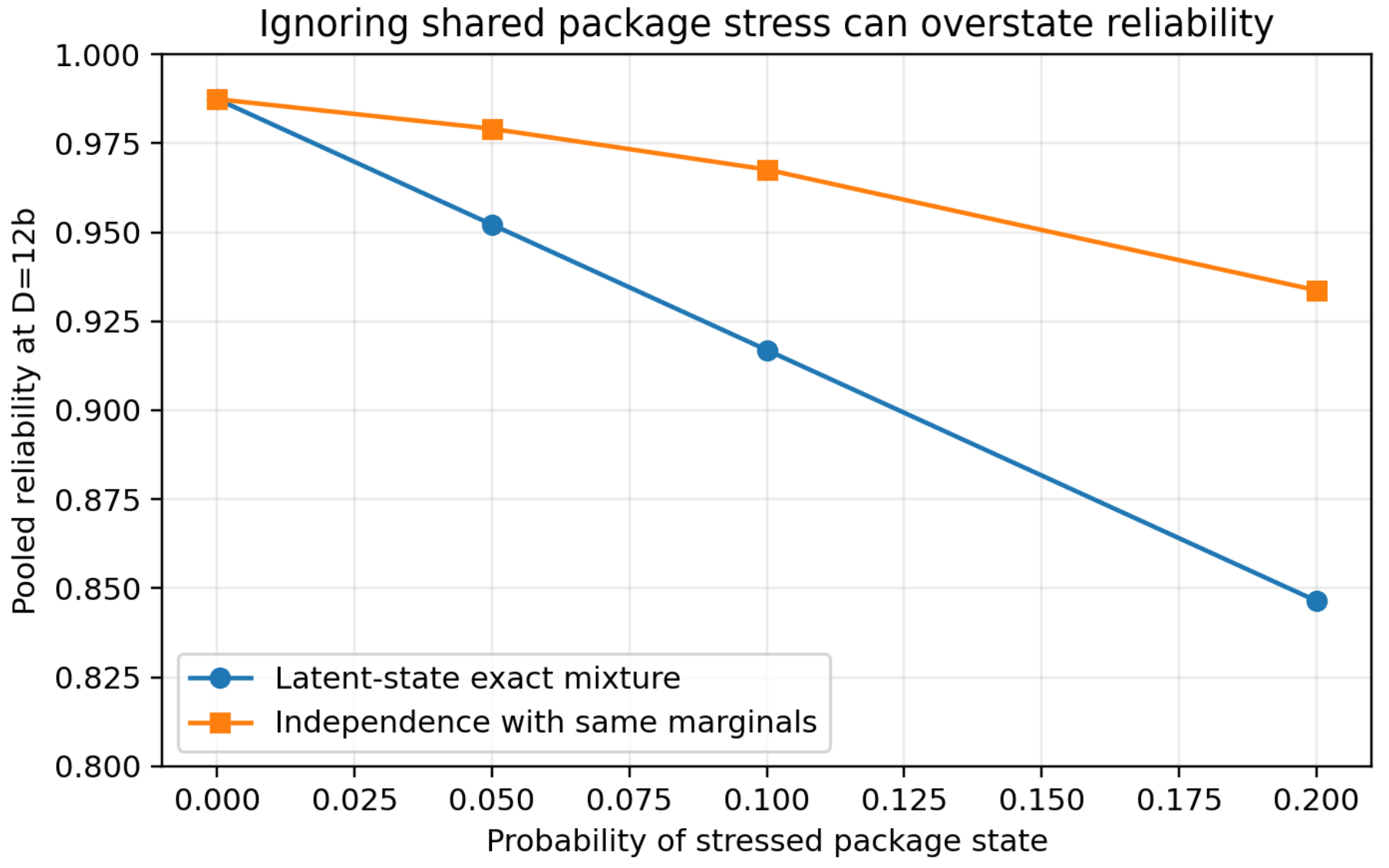

**Figure 4.** Effect of a shared stressed package state. Treating the same unconditional marginals as independent can materially overstate high-demand reliability.

### 6.5. Monte Carlo Sampling Burden

Monte Carlo is included as a stochastic baseline to show the sampling effort required to resolve the high-reliability probabilities computed exactly by the methods of Sections 3 and 4. A high-reliability case: $U = 16$, $(p_{\text{full}}, p_{\text{deg}}, p_{\text{fail}}) = (0.85, 0.10, 0.05)$, $\alpha = 0.5$, $D = 10b$, exact $R_{\text{pool}} = 0.9998267173$. Twenty independent replications at each sample size.

**Table 10.** Monte Carlo sampling burden for $U = 16$, $D = 10b$ and exact $R_{\text{pool}} = 0.9998267173$, over twenty replications at each sample size.

| Samples/run | Mean estimate | Median \|error\| | 95th-percentile \|error\| |
|---|---|---|---|
| 10,000 | 0.9997850 | $7.33 \times 10^{-5}$ | $3.37 \times 10^{-4}$ |
| 100,000 | 0.9998190 | $2.50 \times 10^{-5}$ | $5.67 \times 10^{-5}$ |
| 1,000,000 | 0.9998229 | $9.28 \times 10^{-6}$ | $1.78 \times 10^{-5}$ |

Exact algorithms return the same probability without sampling error. Monte Carlo remains useful for validation but is not competitive for high-precision threshold probabilities.

### 6.6. Heterogeneous Upgrade Selection

Equation (25) is illustrated on the first eight benchmark units specified in the Supplementary Material. Demand is 75% of their maximum capacity, and base reliability is 0.828777. Each unit has one optional action transferring 0.08 probability mass from degraded to full, with normalized costs 1.00, 0.95, 1.10, 0.80, 1.20, 0.90, 1.05, and 0.85.

**Table 11.** Minimum-cost heterogeneous upgrade selection over the first eight benchmark units at $D = 6.012431417 b_0$.

| Target R* | Selected upgraded units | Normalized action cost | Exact $R_{\text{pool}}$ |
|---|---|---|---|
| 0.850 | 1, 4 | 1.80 | 0.852558 |
| 0.870 | 1, 4, 5, 8 | 3.85 | 0.871071 |
| 0.890 | 1, 2, 3, 4, 5, 6, 7, 8 | 7.85 | 0.891114 |

The minimum-cost set depends on both unit-specific states and the system threshold. At $R_{\text{tar}} = 0.85$, units 1 and 4 are selected, even though units 4 and 8 are cheaper together but do not reach the target. The example demonstrates exact multistate reliability in selecting heterogeneous, unit-specific interventions.

**6.7. Central-Demand Scalability**

Central-demand scaling tests whether the storage separation persists as system size grows.

Table 12 shows the central-demand regime at $D = 0.50$ of maximum capacity. The traversal peak grows from 17 to 21 entries while the DP grows from 412 thousand to 29 million states and from 25 MB to 1.63 GB, a divergence of four orders of magnitude where pruning helps least.

**Table 12.** Central demand, $D = 0.50$ of maximum capacity. TP peak states against pruned dynamic-program keys and measured allocation.

| U | TP peak states | DP pruned keys | DP measured bytes | DP median time |
|---|---|---|---|---|
| 14 | 17 | 412,121 | 25.02 MB | 60.9 ms |
| 16 | 19 | 3,468,490 | 194.16 MB | 1,457 ms |
| 18 | 21 | 28,933,671 | 1.628 GB | 17,445 ms |

**6.8. Reserved-Unit Floors and the Third Rule**

The guaranteed-success rule of Section 4.5 is exercised on the fourteen-unit benchmark at $D = 0.65$ of maximum capacity ($9.121829426 b_0$), with uniform floor $0.40 b_0$. Reserved-set size varies from none to all fourteen. Each value was checked against independent sub-stochastic convolution.

The guaranteed remainder uses effective floors $\hat{c}_u$ (Section 2.3) rather than nominal floors. Using effective floors reduces node visits by 2.1% at four reserved units and 21.3% at fourteen; the reductions in Table 13 would be lower if nominal floors were used.

**Table 13.** Effect of the guaranteed-success rule as the number of reserved units varies. $U = 14$, $D = 0.65$ of maximum capacity, floor $0.40b_0$. Peak states are identical for both variants; peak traversal bytes are 576 throughout.

| \|S\| | Exact $R_{\text{res}}$ | Nodes, two rules | Nodes, three rules | Reduction | Peak states |
|---|---|---|---|---|---|
| 0 | 0.982315394994 | 1,139,227 | 1,139,227 | 0.0% | 13 |
| 4 | 0.689450457409 | 510,378 | 412,383 | 19.2% | 13 |
| 7 | 0.533157610196 | 237,261 | 172,708 | 27.2% | 13 |
| 10 | 0.408981300859 | 88,870 | 56,215 | 36.7% | 13 |
| 12 | 0.342467126261 | 39,815 | 20,786 | 47.8% | 12 |
| 14 | 0.286723427738 | 14,771 | 5,273 | 64.3% | 12 |

Reliability falls steeply with reserved-set size, from 0.982315 to 0.286723, demonstrating that pooled capacity overstates deliverable service when units are pinned. The node-visit reduction from the third rule grows from 0% to 64.3%, while peak state count is unchanged (13 falling to 12), confirming that the rule reduces work, not storage. The zero-reservation row is a consistency check: with $\mathcal{R} = \emptyset$, the model reduces exactly to the unreserved problem.

## 7. DISCUSSION AND CONCLUSIONS

Paper 1 [29] derives closed-form results within the binary abstraction. This paper starts where it fails: intermediate bandwidth, heterogeneous units, and multistate capacity distributions. Experiments quantify three distinct modeling errors: binary mapping, pooled-vs-replicated semantics, and common-cause dependence. They also confirm that TP-mBAT is regime-specific, not a universal DP replacement. In the 16-unit commensurate control, DP is superior (273 updates vs. 362,506 node visits). In the 14-unit incommensurate benchmark, the separation is retained state (7 entries vs. 539 states; 17 vs. 412,121 at central demand), as explained by Proposition 1.

The framework complements rather than replaces binary models, reducing to $k$-out-of-$n$ as a special case. Key contributions: TP-mBAT with $O(U \cdot L_{\max})$ traversal storage, Proposition 1 establishing the solver-selection boundary, probability-transfer sensitivity and heterogeneous upgrades, and reserved-unit floors with a third exact pruning rule. Limitations include static probabilities and exponential worst-case complexity; future work will link fixed probabilities to dynamic thermal/degradation occupancy models.

### Data and Code Availability

The analytical formulas and algorithm definitions are contained in the manuscript. The C++ benchmark source, complete worked-example trace, omitted scalability diagnostics, exact heterogeneous benchmark vector, α-sweep calculations, upgrade-selection data, instance specifications, Monte Carlo seeds, and numerical-analysis script are provided as Supplementary Material for reproducibility.